\documentclass[pra,showpacs,showkeys,twocolumn]{revtex4}
\usepackage{graphicx,amsmath,amsfonts,amssymb}
\usepackage{times}
\begin{document}
\title{Photonic two-qubit parity gate with tiny cross-Kerr nonlinearity}
\author{Xin-Wen Wang,$^{1,2,}$\footnote{xwwang@mail.bnu.edu.cn} Deng-Yu Zhang,$^{1,}$\footnote{dyzhang672@163.com}
   Shi-Qing Tang,$^1$ Li-Jun Xie,$^1$ Zhi-Yong Wang,$^3$\footnote{wzyong@cqut.edu.cn} and Le-Man Kuang$^{2,}$\footnote{lmkuang@hunnu.edu.cn}}
 \affiliation{$^1$Department of Physics and Electronic Information Science,
  Hengyang Normal University, Hengyang 421008, People's Republic of China\\
 $^2$Key Laboratory of Low-Dimensional Quantum Structures and Quantum Control of Ministry of Education, and Department of Physics,
  Hunan Normal University, Changsha 410081, People's Republic of China\\
 $^3$School of Optoelectronic Information, Chongqing University of Technology, Chongqing 400054, People's Republic of
 China
 }

\begin{abstract}
The cross-Kerr nonlinearity (XKNL) effect can induce efficient
photon interactions in principle with which photonic multiqubit
gates can be performed using far fewer physical resources than
linear optical schemes. Unfortunately, it is extremely challenging
to generate giant cross-Kerr nonlinearities. In recent years, much
effort has been made to perform multiqubit gates via weak XKNLs.
However, the required nonlinearity strengths are still difficult to
achieve in the experiment. We here propose an XKNL-based scheme for
realizing a two-photon polarization-parity gate, a universal
two-qubit gate, in which the required strength of the nonlinearity
could be orders of magnitude weaker than those required for previous
schemes. The scheme utilizes a ring cavity fed by a coherent state
as a quantum information bus which interacts with a path mode of the
two polarized photons (qubits). The XKNL effect makes the bus pick
up a phase shift dependent on the photon number of the path mode.
Even when the potential phase shifts are very small they can be
effectively measured using photon-number resolving detectors, which
accounts for the fact that our scheme can work in the regime of tiny
XKNL. The measurement outcome reveals the parity (even parity or odd
parity) of the two polarization qubits.

\end{abstract}



\maketitle

\section{Introduction}
The optical system is among the most popular physical systems for
implementing quantum computation. This is mainly due to the facts
that light quantum states are generally more robust against
decoherence than most massive qubit systems, and that the
computation programs can be implemented by simple optical elements
plus photodetections. In addition, all-optical quantum computation
can be combined with quantum communication without qubit
interconversion. In all-optical quantum information processing
(QIP), the qubits are usually encoded by single-photon polarization
states (other types of photonic qubits can be usually converted to
polarization qubits by optical apparatuses). Parametric
down-conversion can produce polarization-entangled photon pairs and
heralded single photons. Beam splitters (BSs) and wave plates can be
used to accomplish arbitrary single-qubit rotations. To complete a
universal gate set for quantum computation, the key point that is
then needed is an appropriate two-qubit quantum gate, such as a
two-qubit parity gate \cite{75PRA032339,76PRA052312} from which a
controlled-NOT (CNOT) gate can be readily constructed
\cite{64PRA062311,93PRL250502,93PRL020501}. Note that a universal
gate set can also serve many quantum communication protocols, in
that it can be utilized to implement their required entangled-state
joint measurements and entangled-channel generations.

To implement a photonic two-qubit gate, nonlinear interactions
between individual photons are required. Linear optical elements
plus photodetections can induce effective nonlinear photon
interactions in principle. This way, however, is nondeterministic,
and needs consuming substantial ancillary photon resources for
achieving a high efficiency
\cite{409N46,95PRL040502,79RMP135,79PRA042326, 82PRA022323}, which
is the main obstacle to large-scale QIP with linear optics. The
required optical nonlinearity can also be achieved directly using a
cross-Kerr medium that can be described by an interaction
Hamiltonian of the form $H=-\hbar\chi a^+_pa_pa^+_sa_s$
\cite{79RMP135,32PRA2287}. Here $a_p$ ($a^+_p$) and $a_s$ ($a^+_s$)
are, respectively, the annihilation (creation) operators of modes
$p$ and $s$, and $\chi$ is the strength of the nonlinearity.
Transforming the mode $p$ ($s$) using this Hamiltonian will induce a
phase shift that depends on the number of photons in the mode $s$
($p$). Indeed, the mode transformations of the two beams are
$a_p\rightarrow a_p\exp(i\theta a^+_sa_s)$ and $a_s\rightarrow
a_s\exp(i\theta a^+_pa_p)$, where $\theta=\chi t$ with $t$ being the
interaction time. When $\theta=\pi$, a two-photon controlled-phase
gate is naturally implemented, from which a CNOT gate can also be
easily constructed. With the giant cross-Kerr phase shift (XKPS),
many schemes for realizing optical quantum nondemolition (QND)
measurements, photonic quantum gates, optical entangled states, and
quantum communication protocols have been proposed (see, e.g.,
\cite{32PRA2287,396N537,62PRA032304,62PRL2124,75PRA034303,51JMO1211,85PRL445}).

 Unfortunately, even the largest natural XKNL is extremely weak. Operating in the
single-photon regime with a mode volume of about 0.1
$\mathrm{cm}^3$, the XKPS is only $\theta\approx 10^{-18}$
\cite{66PRA063814}. This makes cross-Kerr-based optical quantum
gates and QIP extremely challenging. To obtain a giant XKPS, one can
in principle lengthen the cross-Kerr interaction time by
manufacturing a long fiber with the cross-Kerr materials
\cite{3NP95}. In this case, however, photon losses and self-phase
modulation in the cross-Kerr medium will prevent the gate from
operating properly \cite{73PRA052320,77PRA013808}. Since the end of
last century, much effort has been made to generate larger XKNLs
using electromagnetically induced transparency (EIT) (see, e.g.,
\cite{21OL1936,413N273,64PRA023805,91PRL093601,72PRA062319,76PRA052324,101PRL073602,79PRB193404,81PRA053829}).
Although considerable progress has been made, the experimentally
available XKNL still cannot satisfy the requirement \cite{77RMP633}.
Recently, Nemoto and Munro \cite{93PRL250502} proposed a scheme for
realizing a nearly deterministic two-photon parity gate with weak
XKNLs, in which the nonlinearity effect is `amplified' by an intense
coherent state bus. Thereafter, similar schemes for implementing
photonic two-qubit parity and CNOT gates were developed
\cite{83PRA054303,74PRA060302,7NJP137,78PRA022326,79PRA022301}. All
these schemes involve a building block of successive cross-Kerr
interactions between an intense coherent state probe beam---that
acts as a quantum information bus---and a pair of single-photon
qubit beams. To make these schemes work in the regime of
$\theta\thicksim 10^{-2}$, giant intensities of coherent states or
rounds of operations (each round involves two cross-Kerr
interactions) are required. These requirements, however, are very
difficult to achieve in the experiment, or even go beyond the
reasonability. In addition, it is still an experimental challenge to
achieve such a value of XKPS. This idea has also been widely
employed in the research on the Bell-state measurement, generation
of entangled states and coherent state superpositions, and so on
(see, e.g., \cite{71PRA060302,428PR53,79PRA035802,83PRA052303,
84PRA023810}).

In this paper, we propose a scheme for realizing near
deterministically a photonic two-qubit parity gate using a tiny
XKNL. The involved XKNL can be several orders of magnitude weaker
than the aforementioned schemes. The scheme employs an optical
device based on a high quality ring cavity coupled to an external
traveling wave (signal mode) through a cross-Kerr medium
\cite{61PRA053817}. The cavity mode is fed by a coherent state and
serves as a probe, and the detection outcomes at the cavity output
ports reveal the parity (even parity or odd parity) of the two
qubits without destroying the photons.

The paper is organized as follows. In Sec.~II, we introduce the idea
of performing a two-qubit parity gate with a tiny XKNL. In Sec.~III,
we discuss the effects of some nonideal cases on the parity gate:
Sec. III A focuses on the imperfection of the detectors, and Sec.
III B on the photon losses of the coherent state bus. Finally,
concluding remarks are given in Sec.~IV.

\section{Two-qubit parity gate}

The schematic setup of realizing the two-qubit parity gate is
depicted in Fig.~1. For simplicity, we suppose that the two beam
splitters (BSs) have the same transmissivity $\tau$ and their
absorptions are negligible. The cavity is fed by a coherent state
$|\alpha\rangle$ in the input mode $i_1$, whereas the input mode
$i_2$ is left unexcited. After passing through the polarizing beam
splitter (PBS) $PBS_1$, each of the two polarized photons will be in
one of the path modes $s_1$ and $s_2$, which depends on their
polarization states. The mode $s_1$ (signal mode) is coupled to the
cavity mode through the cross-Kerr medium. The two output ports of
the cavity are monitored, respectively, by the photon-number
resolving (PNR) detectors $D_1$ and $D_2$ (when a certain condition
is satisfied, $D_2$ can be omitted as shown later). $PBS_2$ serves
as separating the two photons into different spatial modes.
\begin{figure}
  \center
  \includegraphics{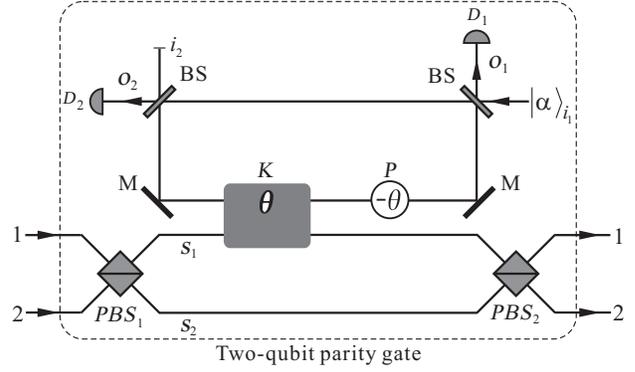}
  \caption{Schematic setup for the two-qubit parity gate. $PBS_1$ and
$PBS_2$ are two polarizing beam splitters, which transmit the
horizontal polarization component and reflect the vertical
component. Two BSs with low transmissivity $\tau$ and two mirrors
(M) constitute a ring cavity which is fed by a coherent state in the
mode $i_1$. $K$ is a cross-Kerr medium by which every photon in the
mode $s_1$ induces a phase shift $\theta$ on the coherent state. $P$
is a static phase shifter that puts a phase shift $-\theta$ on the
coherent state. The cavity transmissivity and reflectivity are
dependent on the phase shift on the coherent state. $D_1$ and $D_2$
are two photon-number resolving detectors, and their detection
outcomes reveal the parity (even parity or odd parity) of the two
qubits without destroying the photons. Note that a simple local
rotation (which depends on the outcomes of $D_1$ and $D_2$) on qubit
1 or 2 via a feedforward process is not shown in the sketch.}
\end{figure}

Before describing the performance of the parity gate in more detail,
we first analyze the dependence of the states of the modes $o_1$ and
$o_2$ on the state of the mode $s_1$. The input-output relations for
the cavity are given by \cite{61PRA053817,40JPB63}
\begin{eqnarray}
\label{transformation}
 && a^+_{o_1}=\kappa_{n}a^{+}_{i_1}+ \exp[i(1-n)\theta]\sigma_{n}a^{+}_{i_2},      \nonumber\\
 && a^+_{o_2}=\sigma_{n}a^{+}_{i_1}+\kappa_{n}a^{+}_{i_2},
\end{eqnarray}
where the cavity transmissivity $\sigma_{n}$ and reflectivity
$\kappa_{n}$ are dependent on the photon number $n$ in the signal
mode $s_1$, both of which read
\begin{eqnarray}
\label{parameter}
  &&   \kappa_{n}=\frac{\sqrt{1-\tau}\left\{\exp[i(1-n)\theta]-1\right\}}{1-(1-\tau)\exp[i(1-n)\theta]},   \nonumber\\
  &&   \sigma_{n}=\frac{\tau}{1-(1-\tau)\exp[i(1-n)\theta]},      \nonumber\\
  &&  |\sigma_{n}|^2= \left[1+4\frac{1-\tau}{\tau^2}\sin^2\frac{(1-n)\theta}{2} \right]^{-1},      \nonumber\\
  &&   |\kappa_{n}|^2=1-|\sigma_{n}|^2.
\end{eqnarray}
Suppose that the mode $s_1$ is initially in an entangled state with
the mode $s_2$,
\begin{equation}
 |\psi\rangle_{s_1s_2}=\sum\limits_{n,m}C_{nm}|n\rangle_{s_1}|m\rangle_{s_2},
\end{equation}
where $|n\rangle_{s_1}$ and $|m\rangle_{s_2}$ denote the
photon-number or Fock states of the modes $s_1$ and $s_2$,
respectively. Then the input state of the total system is
\begin{equation}
 |\psi\rangle_{s_1s_2i_1i_2}=\sum\limits_{n,m}C_{nm}|n\rangle_{s_1}|m\rangle_{s_2}|\alpha\rangle_{i_1}|0\rangle_{i_2}.
\end{equation}
According to Eq.~(\ref{transformation}), we can obtain the output
state of the total system,
\begin{equation}
\label{general}
 |\psi\rangle_{s_1s_2o_1o_2}=\sum\limits_{n,m}C_{nm}|n\rangle_{s_1}|m\rangle_{s_2}|\kappa_n\alpha\rangle_{o_1}|\sigma_n\alpha\rangle_{o_2}.
\end{equation}

Now we consider the two-qubit parity gate. Assume that two
polarization qubits are initially in the state
\begin{eqnarray}
\label{in}
  |\phi^{in}\rangle_{12}&=&x_0|H\rangle_1|H\rangle_2+x_1|H\rangle_1|V\rangle_2\nonumber\\
   && +x_2|V\rangle_1|H\rangle_2 + x_3|V\rangle_1|V\rangle_2,
\end{eqnarray}
where $H$ and $V$ denote the horizontal and vertical polarizations,
respectively. This state may be separable or entangled depending on
whether they have interacted previously. Because the PBSs transmit
the horizontal polarization component and reflect the vertical
component, after passing through $PBS_1$ (see Fig.~1) the incident
state becomes
\begin{eqnarray}
  |\phi\rangle_{s_1s_2}&=&x_0|H\rangle_{s_1}|H\rangle_{s_2}+x_3|V\rangle_{s_1}|V\rangle_{s_2}\nonumber\\
                  && +x_1|0\rangle_{s_1}|HV\rangle_{s_2}+x_2|VH\rangle_{s_1}|0\rangle_{s_2}\nonumber\\
                  &=&x_0|1\rangle_{s_1}|1\rangle_{s_2}+x_3|1\rangle_{s_1}|1\rangle_{s_2}\nonumber\\
                   && +x_1|0\rangle_{s_1}|2\rangle_{s_2}+x_2|2\rangle_{s_1}|0\rangle_{s_2}.
\end{eqnarray}
It can be seen that there is only one photon in each path mode
(called balanced) for the initial state $|H\rangle_1|H\rangle_2$ or
$|V\rangle_1|V\rangle_2$, while there are two photons in one path
mode and none in the other (called bunched) for the initial state
$|H\rangle_1|V\rangle_2$ or $|V\rangle_1|H\rangle_2$. The mode $s_1$
is then coupled to the cavity mode through a cross-Kerr medium, as
shown in Fig.~1. The subsequent $PBS_2$ is used to separate the two
photons into different spatial modes. According to
Eq.~(\ref{general}), the whole system will be finally in the state
\begin{eqnarray}
\label{out}
  |\phi\rangle_{12o_1o_2}&=&\left(x_0|H\rangle_{1}|H\rangle_{2}+x_3|V\rangle_{1}|V\rangle_{2}\right)
      |\kappa_1\alpha\rangle_{o_1}|\sigma_1\alpha\rangle_{o_2}\nonumber\\
    && +x_1|H\rangle_{1}|V\rangle_{2}|\kappa_0\alpha\rangle_{o_1}|\sigma_0\alpha\rangle_{o_2}\nonumber\\
    && +x_2|V\rangle_{1}|H\rangle_{2}|\kappa_2\alpha\rangle_{o_1}|\sigma_2\alpha\rangle_{o_2}.\nonumber\\
\end{eqnarray}
We observe immediately from Eq.~(\ref{parameter}) that $\kappa_1=0$,
$\sigma_1=1$, $\kappa_0=\kappa^*_2=\kappa$, and
$\sigma_0=\sigma^*_2=\sigma$. Note that $\kappa_1=0$ and
$\sigma_1=1$ means the cavity being at resonance and having unit
transmissivity \cite{40JPB63}. Thus the state of Eq.~(\ref{out})
reduces to
\begin{eqnarray}
\label{out1}
  |\phi\rangle_{12o_1o_2}&=&\left(x_0|H\rangle_{1}|H\rangle_{2}+x_3|V\rangle_{1}|V\rangle_{2}\right)
       |0\rangle_{o_1}|\alpha\rangle_{o_2}\nonumber\\
    && +x_1|H\rangle_{1}|V\rangle_{2}|\kappa\alpha\rangle_{o_1}|\sigma\alpha\rangle_{o_2}\nonumber\\
       && +x_2|V\rangle_{1}|H\rangle_{2}|\kappa^*\alpha\rangle_{o_1}|\sigma^*\alpha\rangle_{o_2}.
\end{eqnarray}
For implementing the two-qubit parity gate, that is, obtaining the
even parity state (non-normalized)
\begin{equation}
\label{even}
|\phi^{even}\rangle_{12}=x_0|H\rangle_{1}|H\rangle_{2}+x_3|V\rangle_{1}|V\rangle_{2}
\end{equation}
(two photons have the same polarization and are correlated with each
other) or the odd parity state (non-normalized)
\begin{equation}
\label{odd}
|\phi^{odd}\rangle_{12}=x_1|H\rangle_{1}|V\rangle_{2}+x_2|V\rangle_{1}|H\rangle_{2}
\end{equation}
(two photons have different polarizations and are anti-correlated
with each other), we need to distinguish the probe state
$|0\rangle_{o_1}|\alpha\rangle_{o_2}$ from
$|\kappa\alpha\rangle_{o_1}|\sigma\alpha\rangle_{o_2}$ and
$|\kappa^*\alpha\rangle_{o_1}|\sigma^*\alpha\rangle_{o_2}$, but not
(even in principle) distinguish
$|\kappa\alpha\rangle_{o_1}|\sigma\alpha\rangle_{o_2}$ from
$|\kappa^*\alpha\rangle_{o_1}|\sigma^*\alpha\rangle_{o_2}$. It will
be shown that this task can be accomplished by detecting the photon
numbers of the modes $o_1$ and $o_2$ with the PNR detectors $D_1$
and $D_2$, respectively.

The indistinguishability between $|\kappa\alpha\rangle_{o_1}$ and
$|\kappa^*\alpha\rangle_{o_1}$ with photon-number detection is
evident. The same is true of the indistinguishability between
$|\sigma\alpha\rangle_{o_2}$ and $|\sigma^*\alpha\rangle_{o_2}$. For
distinguishing $|0\rangle_{o_1}|\alpha\rangle_{o_2}$ from
$|\kappa\alpha\rangle_{o_1}|\sigma\alpha\rangle_{o_2}$ and
$|\kappa^*\alpha\rangle_{o_1}|\sigma^*\alpha\rangle_{o_2}$, we in
fact only need to distinguish $|0\rangle_{o_1}$ from
$|\kappa\alpha\rangle_{o_1}$ or $|\kappa^*\alpha\rangle_{o_1}$. It
will be shown later that the photon-number detection on the mode
$o_2$ is just for removing the relative phase shift between the two
components of the odd parity state. The overlap between
$|\kappa\alpha\rangle_{o_1}$ and $|0\rangle_{o_1}$ is very small and
negligible if the amplitude $\kappa\alpha$ is large enough, and they
can be well distinguished from each other. The lower limit of
$|\kappa\alpha|$ depends on the quantum efficiency of the detector
$D_1$ and the allowable error rate. For clarity, we first assume
both $D_1$ and $D_2$ have unity quantum efficiency, and the
imperfect case will be discussed later. Then a measurement on the
mode $o_2$ projects the two qubits and the mode $o_2$ in the state
(non-normalized)
\begin{equation}
\label{e}
|\phi^{e}\rangle\approx\left(x_0|H\rangle_{1}|H\rangle_{2}+x_3|V\rangle_{1}|V\rangle_{2}\right)|\alpha\rangle_{o_2}
\end{equation}
for the photon number  $n_{o_1}=0$, or (non-normalized)
\begin{eqnarray}
\label{o}
|\phi^{o}\rangle&=&x_1 e^{in_{o_1}arg(\kappa)}|H\rangle_{1}|V\rangle_{2}|\sigma\alpha\rangle_{o_2}\nonumber\\
         && +x_2 e^{-in_{o_1}arg(\kappa)}|V\rangle_{1}|H\rangle_{2}|\sigma^*\alpha\rangle_{o_2}\nonumber\\
      &=&\sum\limits_{n_{o_2}=0}^{\infty}f_{n_{o_2}}\left\{x_1 e^{i\varphi(n_{o_1},n_{o_2})}|H\rangle_{1}|V\rangle_{2}\right.\nonumber\\
         && \left.+x_2 e^{-i\varphi(n_{o_1},n_{o_2})}|V\rangle_{1}|H\rangle_{2}\right\}|n_{o_2}\rangle
\end{eqnarray}
for $n_{o_1}>0$, where
$f_{n_{o_2}}=e^{-|\sigma\alpha|^2/2}(|\sigma|\alpha)^{n_{o_2}}/\sqrt{n_{o_2}!}$,
$\varphi(n_{o_1},n_{o_2})=n_{o_1}Arg(\kappa)+n_{o_2}Arg(\sigma)$,
and the identities $arg(\kappa^*)\equiv-Arg(\kappa)$ and
$arg(\sigma^*)\equiv-Arg(\sigma)$ have been utilized. We have used
the approximate equality ($\approx$) in Eq.~(\ref{e}) as there is a
small but finite probability that the state (\ref{o}) can also occur
for $n_{o_1}=0$. If $|x_0|=|x_1|=|x_2|=|x_3|=1/2$ (without loss of
generality, we shall take the same value in the following context),
the probability of this error occurring is given by
\begin{eqnarray}
\label{error}
 P_{err}=\frac{1}{2}|\langle0|\kappa\alpha\rangle|^2=\frac{1}{2}\exp\left(-|\kappa\alpha|^2\right),
\end{eqnarray}
which is less than $10^{-4}$ when $|\kappa\alpha|>3$. Thus,
$|\kappa\alpha\rangle_{o_1}$ and $|0\rangle_{o_1}$ can be
discriminated near deterministically. Experimental implementations
of discriminating between a coherent state and a vacuum state using
the photon-number measurement have been recently reported
\cite{446N774,104PRL100505}.

It can be seen that for the measurement outcome $n_{o_1}=0$ the mode
$o_2$ is disentangled and the two qubits are directly projected in
the even parity state of Eq.~(\ref{even}). However, for $n_{o_1}>0$
the two qubits are still entangled with the mode $o_2$. To
disentangle the mode $o_2$, one needs to perform another
photon-number measurement on $o_2$ and obtain its photon number
$n_{o_2}$. Then one obtains an odd parity state with the two
components picking up an unwanted relative phase shift
$2\varphi(n_{o_1},n_{o_2})$. The phase shift
$2\varphi(n_{o_1},n_{o_2})$ can be eliminated via a classical
feedforward operation (In many computational circuits the
phase-shift removing operations can be delayed and performed at the
final measurement stage for the qubits). After all the operations
discussed above, we can conclude that a two-qubit parity gate is
accomplished with near one probability.

As shown above, detecting the mode $o_1$ is sufficient for
discriminating between the odd and even parities of the two qubits,
and detecting $o_2$ is just for removing the relative phase shift of
the two odd parity components $|H\rangle_{1}|V\rangle_{2}$ and
$|V\rangle_{1}|H\rangle_{2}$. When $|\sigma\alpha|$ is very small,
the mean photon number ($\bar{n}_{o_2}$) of the states
$|\sigma\alpha\rangle$ and $|\sigma^*\alpha\rangle$ is close to
zero. Then one may omit the detection of $o_2$. This, however, will
yield a small error probability given by
$1-|\langle0|\sigma\alpha\rangle|^2=1-\exp\left(-\bar{n}_{o_2}^2\right)$
which is less than $10^{-3}$ when $\bar{n}_{o_2}<0.001$.

In what follows, we analyze how large XKPS ($\theta$) is required
for realizing the aforementioned parity gate. For a certain value of
$P_{err}$, the value of $\theta$ evidently depends on the values of
$|\alpha|$ and $\tau$. What we are interested in is the weak
nonlinearity regime (i.e., $\theta\ll 1$). In addition, we assume
the transmissivity ($\tau$) of the BSs is also very small. Then,
$|\kappa|~(=|\kappa_0|=|\kappa_2|)$ and
$|\sigma|~(=|\sigma_0|=|\sigma_2|)$ can be approximated as
\begin{eqnarray}
\label{approximate}
  |\kappa|^2\approx\frac{r^2}{r^2+1},\nonumber\\
  |\sigma|^2\approx\frac{1}{r^2+1},
\end{eqnarray}
where $r=\theta/\tau$. According to Eqs.~(\ref{error}) and
(\ref{approximate}), the relationship of $r$ and $|\alpha|$ is given
by
\begin{equation}
 |\alpha|\approx\frac{1}{r}\sqrt{(r^2+1)\ln\frac{1}{ 2P_{err}}}.
\end{equation}
Evidently, $r$ decreases (i.e., $\theta$ decreases for a given
$\tau$) as $|\alpha|$ increases for a certain value of $P_{err}$,
and vice versa. When $P_{err}$ and $|\alpha|$ are given (i.e., $r$
is given), $\theta$ is in inverse proportion to $1/\tau$. These
results imply that our scheme can work in the regime of tiny XKNL.

We take $P_{err}=10^{-4}$ as an example. Then the dependence
relation of $r$ and $|\alpha|$ is shown in Fig.~2. It can be seen
that $|\alpha|$ slowly increases as $r$ rapidly decreases in the
range $r\gtrsim 2$, while $|\alpha|$ rapidly increases as $r$ slowly
decreases in the range $r\lesssim0.5$. This indicates that a large
decrease in $\theta$ only needs a small increase in $|\alpha|$ for
the case $\theta\gtrsim 2\tau$, while a small decrease in $\theta$
needs a large increase in $|\alpha|$ for the case
$\theta\lesssim\tau/2$. Thus there is a trade-off between $\theta$
and $|\alpha|$. In the following context, we focus on an example
case, $\theta=\tau$. Then the lower $\tau$ is, the smaller regime of
$\theta$ our scheme can work in. For example, when $\tau$ is of the
order of $\{10^{-6},10^{-5},10^{-4}\}$ (these values are available
under current technology \cite{61PRA053817,5NP628}), $\theta$ is
correspondingly of the order of $\{10^{-6},10^{-5},10^{-4}\}$. It
seems that the previous schemes can also work in the regime of these
orders of magnitude of XKPS by increasing the amplitude of the probe
coherent state or the number of round. However, the intensity of the
coherent state or the number of round would go beyond the accessible
or even reasonable values. For example, to make the parity gate
operate properly even for $\theta\sim 10^{-2}$ (assuming the error
probability is also of the order of $10^{-4}$), the amplitude of the
probe coherent state should be $\sim 10^4$ in
Refs.~\cite{93PRL250502,83PRA054303,79PRA022301} (note that a much
more intense ancillary coherent state beam is also required for
accomplishing the homodyne measurement on the probe coherent state
beam \cite{Scully}), and the number of round (each round involves
two cross-Kerr interactions) should also be $\sim 10^4$ in
Ref.~\cite{74PRA060302}. Although the amplitude of the probe
coherent state can be reduced to a certain extent by using
photon-number measurement than homodyne measurement
\cite{7NJP137,74PRA060302,78PRA022326}, inaccessibly or even
unreasonably intense ancillary coherent state would be required for
accomplishing an appropriate displacement on the probe coherent
state. Thus we conclude that our scheme can work in the regime of a
several orders of magnitude weaker XKNL than those schemes mentioned
above.
\begin{figure}
  \center
  \includegraphics[width=9cm,height=6cm]{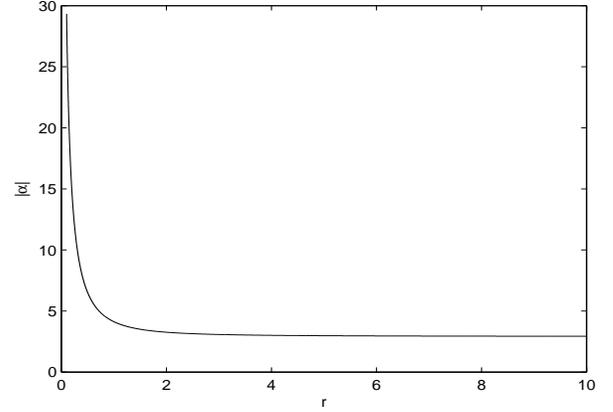}
  \caption{The dependence relation of $|\alpha|$ and $r$, with the
error probability $P_{err}=10^{-4}$.}
\end{figure}

\section{Effects of nonideal cases on the two-qubit parity gate}

\subsection{Imperfections of detectors}

We now consider that the two detectors $D_1$ and $D_2$ have nonunit
quantum efficiencies $\eta_1$ and $\eta_2$, respectively. Then the
error probability (\ref{error}) is replaced by
\begin{eqnarray}
 P^{(\eta_1)}_{err}&=&\frac{1}{2}\mathrm{Tr}\left[|\kappa\alpha\rangle\langle\kappa\alpha|\sum\limits_{n=0}^{\infty}(1-\eta_1)^n|n\rangle\langle
            n|\right]\nonumber\\
   &=&\frac{1}{2}\exp\left[-\eta_1|\kappa\alpha|^2\right].
\end{eqnarray}
Evidently, the imperfection of $D_1$ can be well compensated by
slightly increasing $|\alpha|$. For the same $\kappa$,
$P^{(\eta_1)}_{err}(\alpha')=P_{err}(\alpha)$ with
$\alpha'=\alpha/\sqrt{\eta_1}$, and when, for example, $\eta_1=0.9$,
$|\alpha'|$ is about 4.22 for $|\alpha|=4$ and about 30.57 for
$|\alpha|=29$.

For eliminating the relative phase shift of the two odd parity
components, one needs to know the correct photon numbers of the
modes $o_1$ and $o_2$, as shown before. The imperfections of $D_1$
and $D_2$ will cause an error probability given by
\begin{eqnarray}
\label{suberror}
  P'^{(\eta_1+\eta_2)}_{err}&=&1-\sum\limits_{n=1}^{\infty}|\langle n|\kappa\alpha\rangle|^2\eta_1^n\sum\limits_{m=1}^{\infty}|\langle m|\sigma\alpha\rangle|^2\eta_2^m\nonumber\\
   &=&1-\exp\left[(\eta_1|\kappa|^2+\eta_2|\sigma|^2-1)|\alpha|^2\right].
\end{eqnarray}
For simplicity, we suppose $\eta_1=\eta_2=\eta$. Then the above
equation reduces to
\begin{eqnarray}
\label{suberror}
  P'^{(\eta)}_{err} =1-\exp\left[(\eta-1)|\alpha|^2\right],
\end{eqnarray}
which is independent of the XKPS $\theta$. $P'^{(\eta)}_{err}$ is
small if and only if $(1-\eta)|\alpha|^2$ is small. When
$(1-\eta)|\alpha|^2<0.05$, $P'^{(\eta)}_{err}$ is less than $0.05$.
In addition, for a given value of $\eta$, the larger $|\alpha|$ is,
the larger $P'^{(\eta)}_{err}$ becomes. Thus, for making
$P'^{(\eta)}_{err}$ be very small, the quantum efficiency is
required to be very high when $|\alpha|$ is not very small.
Fortunately, recent reports \cite{71PRA061803,97APL031102,1108.5299}
indicated that $\eta$ could be close to unity with the developing
techniques. We notice that potential dark counts of the detectors
could also cause errors. However, recent experiments
\cite{71PRA061803,97APL031102,1108.5299} demonstrated that
near-unity-efficiency PNR detectors with negligible dark-count rates
could be produced. Note that other schemes
\cite{93PRL250502,83PRA054303,74PRA060302,7NJP137,78PRA022326,79PRA022301}
may be more fragile to the imperfection of the detectors, because
they involve photon-number measurements on very intense coherent
states as mentioned before.

It is worth stressing that $P'^{(\eta_1+\eta_2)}_{err}$ is not the
error probability of distinguishing between the odd and even
parities of the two qubits but the error probability of removing the
relative phase shift of the two odd parity components. The error
probability ($P^{(\eta_1)}_{err}$) of distinguishing the odd parity
components $|HV\rangle$ and $|VH\rangle$ from the even parity
components $|HH\rangle$ and $|VV\rangle$ could be close to zero even
when the detection efficiency is low (in the following context, we
shall use the abbreviation $|H\rangle|V\rangle=|HV\rangle$).
Therefore, even with low-efficiency detectors, our scheme could be
used to implement the Bell-state measurement with near unity
probability \cite{7NJP137,83PRA054303,82PRA032318} and generate
cluster states with a certain probability
\cite{83PRA054303,75PRA042323} as well as serve all other quantum
tasks that involve two-photon polarization-parity detections.

It has been mentioned before that when $|\sigma\alpha|$ is very
small (e.g., $r=100$ and $|\alpha|$ is not too large), the
measurement on the mode $o_2$ may well be omitted. Then the error
probability $P'^{(\eta_1+\eta_2)}_{err}$ vanishes.

\subsection{Photon losses in the bus}

In this section, we discuss the decoherence effect due to photon
absorption in the cross-Kerr medium. Photon losses may occur in both
the bus and the qubit modes. However, photon losses in the coherent
state field is more easier to occur. Therefore, photon losses of the
coherent state field should be the main source of decoherence in the
two-qubit output state when the interaction time is very short
\cite{7NJP137,73PRA052320,79PRA035802}. In what follows, we shall
consider the photon losses in the bus. Such photon losses can be
modeled via a beam splitter of transmissivity $\lambda$ which
discards a portion of the coherent state beam
\cite{7NJP137,64PRA022313}. It is assumed that $\lambda$ does not
vary with time and can be measured in advance through suitable test
experiments \cite{7NJP137,73PRA052320}, so its value is known. Then
Fig.~1 can be altered phenomenally to Fig.~3,
\begin{figure}
  \center
  \includegraphics{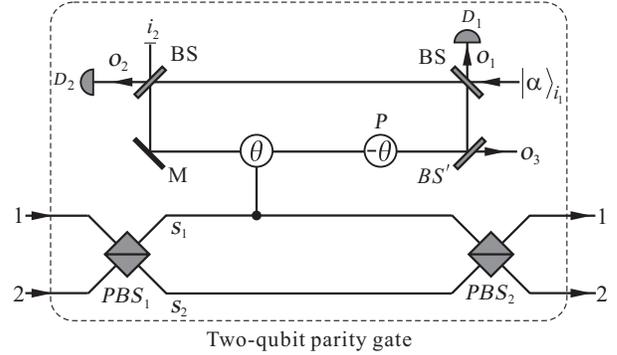}
  \caption{Diagrammatic sketch of the two-qubit parity gate. Each
photon in the mode $s_1$ induces a phase shift $\theta$ on the
coherent state through the cross-Kerr medium. The additional beam
splitter $BS'$ (with transmissivity $\lambda$) is used to model the
photon losses in the coherent state bus.}
\end{figure}
where the beam splitter $BS'$ (with transmissivity $\lambda$) is to
model the photon losses in the cavity field. Following the method of
the authors of \cite{61PRA053817,40JPB63}, we obtain the
input-output relations for the cavity,
\begin{eqnarray}
\label{threeport}
&& a^+_{o_1}=A_na^+_{i_1}+\frac{\tau\sqrt{1-\lambda}e^{i(1-n)\theta}}{\Gamma}a^+_{i_2}+\frac{\sqrt{\tau\lambda}}{\Gamma}a^+_{i_3},\nonumber\\
&&a^+_{o_2}=B_na^+_{i_1}+A_na^+_{i_2}+\frac{\sqrt{\lambda\tau(1-\tau)}}{\Gamma}a^+_{i_3},\nonumber\\
&&a^+_{o_3}=C_na^+_{i_1}+\frac{C_n}{\sqrt{1-\tau}}a^+_{i_2}+\frac{\rho
e^{i(1-n)\theta}-\sqrt{1-\lambda}}{\Gamma}a^+_{i_3},\nonumber\\
\end{eqnarray}
where
\begin{eqnarray}
\label{ABC}
&&\Gamma=1-(1-\tau)\sqrt{1-\lambda}e^{i(1-n)\theta},\nonumber\\
 &&A_n=\frac{\sqrt{1-\tau}\left[\sqrt{1-\lambda}e^{i(1-n)\theta}-1\right]}{\Gamma},\nonumber\\
&&B_n=\frac{\tau}{\Gamma},\nonumber\\
&&C_n=\frac{\sqrt{\lambda\tau(1-\tau)}e^{i(1-n)\theta}}{\Gamma},
\end{eqnarray}
and $n$ ($=0,1,2$) is the photon number in the mode $s_1$. The mode
$o_3$ that denotes the photon losses in the cavity field has to be
traced out \cite{7NJP137,64PRA022313}. Then, the output state of
Eq.~(\ref{out1}) is replaced by
\begin{widetext}
\begin{eqnarray}
\label{out2}
\rho_{12o_1o_2}&=&|\phi^{even}\rangle_{12}\langle\phi^{even}|\otimes|A_1\alpha\rangle_{o_1}\langle A_1\alpha|\otimes|B_1\alpha\rangle_{o_2}\langle B_1\alpha|\nonumber\\
&&+|x_1|^2|HV\rangle_{12}\langle HV|\otimes|A_0\alpha\rangle_{o_1}\langle A_0\alpha|\otimes|B_0\alpha\rangle_{o_2}\langle B_0\alpha|\nonumber\\
&&+|x_2|^2|VH\rangle_{12}\langle VH|\otimes|A^*_0\alpha\rangle_{o_1}\langle A^*_0\alpha|\otimes|B^*_0\alpha\rangle_{o_2}\langle B^*_0\alpha|\nonumber\\
&&+x^*_1y_1|\phi^{even}\rangle_{12}\langle HV|\otimes|A_1\alpha\rangle_{o_1}\langle A_0\alpha|\otimes|B_1\alpha\rangle_{o_2}\langle B_0\alpha|\nonumber\\
&&+x_1y_1|HV\rangle_{12}\langle\phi^{even}|\otimes|A_0\alpha\rangle_{o_1}\langle A_1\alpha|\otimes|B_0\alpha\rangle_{o_2}\langle B_1\alpha|\nonumber\\
&&+x^*_2y_2|\phi^{even}\rangle_{12}\langle VH|\otimes|A_1\alpha\rangle_{o_1}\langle A^*_0\alpha|\otimes|B_1\alpha\rangle_{o_2}\langle B^*_0\alpha|\nonumber\\
&&+x_2y_2|VH\rangle_{12}\langle\phi^{even}|\otimes|A^*_0\alpha\rangle_{o_1}\langle A_1\alpha|\otimes|B^*_0\alpha\rangle_{o_2}\langle B_1\alpha|\nonumber\\
&&+x_1x^*_2y_3|HV\rangle_{12}\langle VH|\otimes|A_0\alpha\rangle_{o_1}\langle A^*_0\alpha|\otimes|B_0\alpha\rangle_{o_2}\langle B^*_0\alpha|\nonumber\\
&&+x^*_1x_2y_3|VH\rangle_{12}\langle
HV|\otimes|A^*_0\alpha\rangle_{o_1}\langle
A_0\alpha|\otimes|B^*_0\alpha\rangle_{o_2}\langle B_0\alpha|,
\end{eqnarray}
where
\begin{eqnarray}
y_1=|\langle C_0\alpha|C_1\alpha\rangle|^2,~~~~ y_2=|\langle
C^*_0\alpha|C_1\alpha\rangle|^2,~~~ y_3=|\langle
C^*_0\alpha|C_0\alpha\rangle|^2,
\end{eqnarray}
and the relations $A_2=A^*_0$ and $B_2=B^*_0$ have been utilized. In
this case, the probe mode $o_1$ must be displaced by an amount
$D(-A_1\alpha)=\exp\left(A_1\alpha^*a_{o_1}-A_1\alpha a^+_{o_1}
\right)$ ($A_1^*=A_1$) prior to the measurement. Then the state of
Eq.~(\ref{out2}) evolves to
\begin{eqnarray}
\label{out3}
\rho_{12o_1o_2}&=&|\phi^{even}\rangle_{12}\langle\phi^{even}|\otimes|0\rangle_{o_1}\langle 0|\otimes|B_1\alpha\rangle_{o_2}\langle B_1\alpha|\nonumber\\
&&+|x_1|^2|HV\rangle_{12}\langle HV|\otimes|(A_0-A_1)\alpha\rangle_{o_1}\langle (A_0-A_1)\alpha|\otimes|B_0\alpha\rangle_{o_2}\langle B_0\alpha|\nonumber\\
&&+|x_2|^2|VH\rangle_{12}\langle VH|\otimes|(A_0-A_1)^*\alpha\rangle_{o_1}\langle (A_0-A_1)^*\alpha|\otimes|B^*_0\alpha\rangle_{o_2}\langle B^*_0\alpha|\nonumber\\
&&+x^*_1y_1e^{i|\alpha|^2A_1\mathrm{Im}A_0}|\phi^{even}\rangle_{12}\langle HV|\otimes|0\rangle_{o_1}\langle (A_0-A_1)\alpha|\otimes|B_1\alpha\rangle_{o_2}\langle B_0\alpha|\nonumber\\
&&+x_1y_1e^{-i|\alpha|^2A_1\mathrm{Im}A_0}|HV\rangle_{12}\langle\phi^{even}|\otimes|(A_0-A_1)\alpha\rangle_{o_1}\langle 0|\otimes|B_0\alpha\rangle_{o_2}\langle B_1\alpha|\nonumber\\
&&+x^*_2y_2e^{i|\alpha|^2A_1\mathrm{Im}A^*_0}|\phi^{even}\rangle_{12}\langle VH|\otimes|0\rangle_{o_1}\langle (A_0-A_1)^*\alpha|\otimes|B_1\alpha\rangle_{o_2}\langle B^*_0\alpha|\nonumber\\
&&+x_2y_2e^{-i|\alpha|^2A_1\mathrm{Im}A^*_0}|VH\rangle_{12}\langle\phi^{even}|\otimes|(A_0-A_1)^*\alpha\rangle_{o_1}\langle 0|\otimes|B^*_0\alpha\rangle_{o_2}\langle B_1\alpha|\nonumber\\
&&+x_1x^*_2y_3e^{-i2|\alpha|^2A_1\mathrm{Im}A_0}|HV\rangle_{12}\langle VH|\otimes|(A_0-A_1)\alpha\rangle_{o_1}\langle (A_0-A_1)^*\alpha|\otimes|B_0\alpha\rangle_{o_2}\langle B^*_0\alpha|\nonumber\\
&&+x^*_1x_2y_3e^{i2|\alpha|^2A_1\mathrm{Im}A_0}|VH\rangle_{12}\langle
HV|\otimes|(A_0-A_1)^*\alpha\rangle_{o_2}\langle
(A_0-A_1)\alpha|\otimes|B^*_0\alpha\rangle_{o_2}\langle B_0\alpha|.
\end{eqnarray}
Note that the amplitudes of $|HV\rangle_{12}$ and $|VH\rangle_{12}$
have picked up a phase shift due to the displacement. These phase
shifts are unwanted but can be simply removed by static phase
shifters (no feedforward is required). After the operations as
mentioned above (performing photon-number measurements on the modes
$o_1$ and $o_2$, and eliminating the unwanted phase shifts via a
classical feedforward process), the two qubits end in the even
parity state $|\phi^{even}\rangle_{12}$ for $n_{o_1}=0$ or an odd
parity state
\begin{eqnarray}
\rho^{odd}_{12}&=&|x_1|^2|HV\rangle_{12}\langle
HV|+|x_2|^2|VH\rangle_{12}\langle VH|
 +x_1x_2^*y_3|HV\rangle_{12}\langle
VH|+x^*_1x_2y_3|VH\rangle_{12}\langle HV|
\end{eqnarray}
for $n_{o_1}>0$. We have assumed that the detectors are perfect.
Obviously, the potentially obtained even parity state is exactly the
target state $|\phi^{even}\rangle_{12}$ as given in
Eq.~(\ref{even}), while the possibly obtained odd parity state is a
mixed state which is different from the desired state
$|\phi^{odd}\rangle_{12}$ as shown in Eq.~(\ref{odd}). This
indicates that the photon losses in the bus only cause decoherence
for the odd parity state of the two qubits.

Let $|x_0|=|x_1|=|x_2|=|x_3|=1/2$. Then the error probability of
distinguishing the even parity components $|HH\rangle$ and
$|VV\rangle$ from the odd parity components $|HV\rangle$ and
$|VH\rangle$ is
\begin{eqnarray}
P_{e}&=&\frac{1}{2}|\langle 0|(A_0-A_1)\alpha\rangle|^2\nonumber\\
&=&
\frac{1}{2}\exp\left\{\frac{2\tau^2(1-\tau)(1-\lambda)(\cos\theta-1)|\alpha|^2}{\left[1-2(1-\tau)\sqrt{1-\lambda}\cos\theta+(1-\tau)^2(1-\lambda)\right]
\left[1-(1-\tau)\sqrt{1-\lambda}\right]^2}\right\}.
\end{eqnarray}
Figure 4 shows that $P_{e}$ only depends on the ratio of $\lambda$
(photon loss parameter) to $\theta$ for a given $|\alpha|$ and
$\theta=\tau$, and it universally decreases with the increase of
$|\alpha|$. In a word, $P_{e}$ is approximate to zero and negligible
when $\lambda$ does not exceed a certain threshold value depending
in a nontrivial way upon the values of $\theta$ and $|\alpha|$. The
overlap between the potentially obtained odd parity state and the
desired odd parity state $|\phi^{odd}\rangle_{12}$ is given by
\begin{eqnarray}
F_{odd}&=&\frac{_{12}\langle\phi^{odd}|\rho^{odd}_{12}|\phi^{odd}\rangle_{12}}{\mathrm{tr}\left(|\phi^{odd}\rangle_{12}\langle\phi^{odd}|\right)\mathrm{tr}(\rho^{odd}_{12})}\nonumber\\
&=&
\frac{1}{2}+\frac{1}{2}\exp\left\{\frac{-4\tau(1-\tau)\lambda|\alpha|^2\sin^2\theta}{\left[1-2(1-\tau)\sqrt{1-\lambda}\cos\theta+(1-\tau)^2(1-\lambda)\right]^2}\right\}.
\end{eqnarray}
\end{widetext}
\begin{figure}
  \center
  \includegraphics[width=11cm,height=7cm]{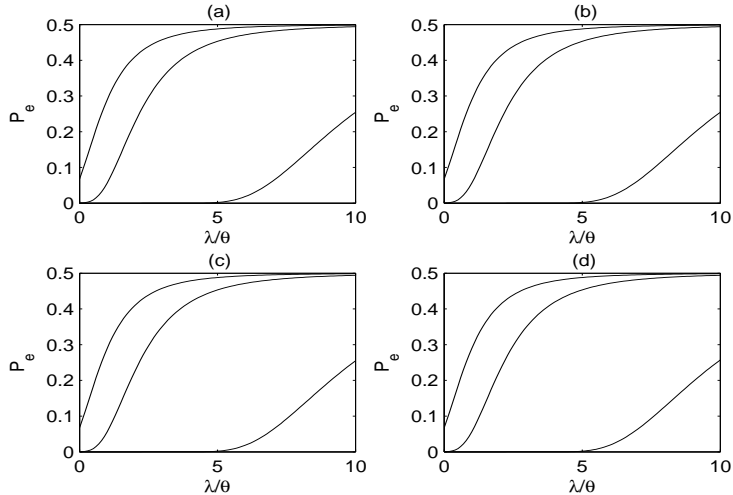}
  \caption{The error probability $P_e$ against the ratio of $\lambda$
(photon loss parameter) to $\theta$. $\tau$ is equal to $\theta$.
From top to bottom in each graph, the curves correspond to
$|\alpha|$=2, 4, and 30, respectively. (a) $\theta=10^{-6}$. (b)
$\theta=10^{-5}$. (c) $\theta=10^{-4}$. (d) $\theta=10^{-3}$.}
\end{figure}
\begin{figure}
  \center
  \includegraphics[width=11cm,height=7cm]{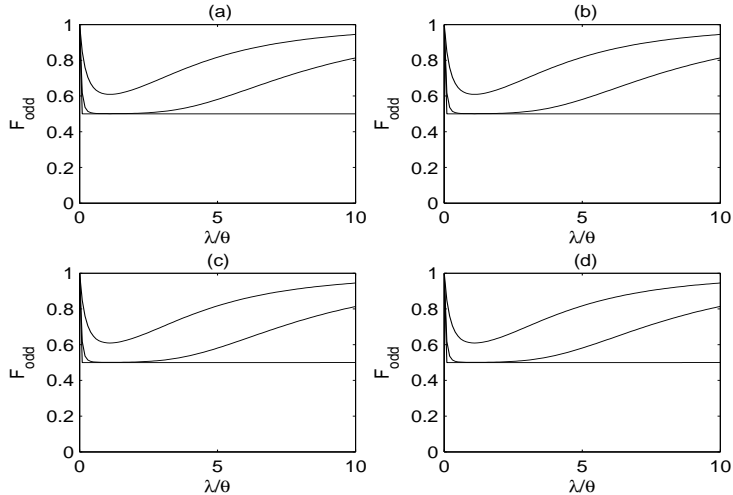}
  \caption{The fidelity $F_{odd}$ against $\lambda/\theta$. $\tau$ is
equal to $\theta$. From top to bottom in each graph, the curves
correspond to $|\alpha|$=2, 4, and 30, respectively. (a)
$\theta=10^{-6}$. (b) $\theta=10^{-5}$. (c) $\theta=10^{-4}$. (d)
$\theta=10^{-3}$.}
\end{figure}
Figure 5 shows the dependence of $F_{odd}$ on $\lambda/\theta$ for a
given $|\alpha|$ and $\theta=\tau$. It can be observed from figures
4 and 5 that the conditions of both $P_{e}$ being close to zero and
$F_{odd}$ being very large cannot be simultaneously satisfied when
$\lambda/\theta$ is not sufficiently small.

As shown above, even when $F_{odd}$ holds small values, the even and
odd parity components could be near deterministically distinguished.
Similar to the foregoing case, our scheme with photon losses in the
bus could be also used to implement the Bell-state measurement with
near unity probability \cite{7NJP137,83PRA054303,82PRA032318} and
generate cluster states with a certain probability
\cite{83PRA054303,75PRA042323} as well as serve all other quantum
tasks that involve two-photon polarization-parity detections.
Finally, we should point out that photon absorption in the
cross-Kerr interactions could be nearly eliminated under certain
conditions by using EIT materials \cite{77PRA033843}.

Note that other potential factors should be also considered for
practical implementation of our scheme, such as self-phase
modulation and non-instantaneous response in the cross-Kerr medium.
Fortunately, these effects may well be canceled through replacing
fibers with EIT materials \cite{77PRA013808,83PRA053826}. In
addition, the spectral effect can be also circumvented under certain
conditions \cite{83PRA053826,9NJP16,0810.2828v2}, and other
resources of error can be dealt with using the standard techniques
available for linear optical quantum computation \cite{79RMP135}.

\section{Concluding remarks}

In conclusion, we have proposed a scheme for realizing a nearly
deterministic two-photon polarization-parity gate using a tiny XKNL.
Like previous XKNL-based schemes, our scheme only needs far fewer
physical resources than linear optical schemes. The scheme employs
an optical device based on a high quality ring cavity constructed by
both two BSs and mirrors and fed by a coherent state. Such a device
can substantially `amplify' the nonlinearity effect, which makes it
possible that our scheme work in the regime of a several orders of
magnitude weaker XKNL than previous schemes. As a consequence, our
scheme is more practical under current XKNL techniques. The
presented two-qubit parity gate plus single-qubit rotations can
constitute a universal gate set for economical and feasible
all-optical quantum computation. Similarly, the two-qubit parity
gate could serve quantum communication systems as it can be used to
realize complete Bell-state measurements
\cite{7NJP137,83PRA054303,82PRA032318}, multiphoton-entanglement
generations \cite{83PRA054303,75PRA042323}, optimal nonlocal
multiphoton-entanglement concentrations \cite{85PRA022311}, and so
on. In addition, we showed that when the quantum efficiencies of the
PNR detectors are less than one and there are photon losses in the
bus, the even and odd parity states of the two polarization qubits
can be near deterministically distinguished. These findings indicate
that even in the nonideal cases, our scheme could efficiently serve
the quantum tasks that involve two-photon polarization-parity
detections.

\begin{acknowledgements}
This work was supported by the National Natural Science Foundation
of China (Grants No. 11004050 and 11075050), the Program for
Changjiang Scholars and Innovative Research Team in University
(Grant No. IRT0964), the Key Project of Chinese Ministry of
Education (Grant No. 211119), the Scientific Research Fund of Hunan
Provincial Education Department (Grants No. 09A013 and 10B013), and
the construct program of the key discipline in Hunan province.
\end{acknowledgements}

\end{document}